# Financial Risks and the Pension Protection Fund:

# Can it Survive Them?

David Blake, John Cotter & Kevin Dowd[*]

November 2006

Abstract

This paper discusses the financial risks faced by the UK Pension Protection Fund (PPF) and what, if anything, it can do about them. It draws lessons from the regulatory regimes under which other financial institutions, such as banks and insurance companies, operate and asks why pension funds are treated differently. It also reviews the experience with other government-sponsored insurance schemes, such as the US Pension Benefit Guaranty Corporation, upon which the PPF is modelled. We conclude that the PPF will live under the permanent risk of insolvency as a consequence of the moral hazard, adverse selection, and, especially, systemic risks that it faces.

[*] David Blake (corresponding author) is the director of the Pensions Institute, Cass Business School, 106 Bunhill Row, London EC1Y 8TZ, UK. Email: d.blake@city.ac.uk. John Cotter is the director of the Centre for Financial Markets, Smurfit School of Business, University College Dublin, Carysfort Avenue, Blackrock, Co. Dublin, Ireland; email: john.cotter@ucd.ie. Kevin Dowd is at the Centre for Risk and Insurance Studies, Nottingham University Business School, Jubilee Campus, Nottingham NG8 1BB, UK; email: Kevin.Dowd@nottingham.ac.uk. The authors are grateful to Victor Dowd for advice in preparing this paper.

# 1 Introduction

The Pension Protection Fund (PPF) is a UK Government-sponsored insurance scheme established by the 2004 Pensions Act with the aim of protecting members of private sector defined benefit (DB) schemes whose firms become insolvent and have insufficient funds in their pension scheme: this is to reassure them that they will still receive most of the pension benefits that they were expecting. The PPF came into operation in April 2005. Participation is mandatory, and pension funds are taken over by the PPF when the sponsoring employer has become insolvent and the pension scheme has insufficient assets to buy out the PPF level of benefits with a life company. Once in the PPF, the scheme can never leave.

The PPF protects 100% of the pension for members above scheme pension age, and 90% of the promised pension for members below scheme pension age (up to a maximum of £25,000 at age 65) using a mixture of scheme individual rates and standardised rules. Pensions in payment are subject to limited priced indexation (LPI) at 2.5%, while deferred pensions are subject to LPI at 5%. Survivors' benefits are also protected.

The compensation is funded by taking on the assets of insolvent schemes and by charging a levy on schemes. The levy is charged to all private sector DB and hybrid occupational pension schemes and is collected by the Pensions Regulator. The levy has three components:

- Pension Protection Levy
    - a 'scheme factors' element which depends on the number of members and the balance between active and retired members
    - a 'risk factors' element (at least 80% of the total charge, although not raised in the first year of operation) which is linked to such factors as the level of underfunding, investment strategy and the sponsor's credit rating.
- Administration Levy, covering set-up and ongoing costs of the PPF.
- Fraud Compensation Levy.

The governance and management of the PPF are in the hands of a Board which is responsible for paying pension compensation, paying fraud compensation, determining the three levies, setting investment strategy, and appointing at least two independent fund managers.

The Government has made it clear that it will not underwrite the PPF. Instead its stated objective is that the PPF must survive on the basis of its powers to set levies and determine its own liabilities. In this paper, we consider the feasibility of this objective (section 6). In order to do this, we need to assess the financial risks that the PPF faces (section 5). But before doing that, it will be instructive to consider how other financial institutions, such as banks and insurance companies, are regulated and deal with the financial risks they face (section 2), what makes pension funds different from these other institutions (section 3), and how other government-sponsored insurance schemes deal with the risks that they face (section 4).

## 2 The Financial Regulation of Banks and Life Assurance Companies

### 2.1 Banks

One of the most important trends in financial regulation is the move towards a common risk-based framework of regulatory capital requirements for financial institutions. For example, the UK financial regulator, the Financial Services Authority (FSA), currently has separate sets of financial regulations (prudential sourcebooks) for banks, building societies, friendly societies, insurers and investment firms, but its ultimate objective is to harmonise these regulations and introduce an integrated prudential sourcebook (PSB) based principally on that for banks. Similar convergence processes are occurring throughout the EU. Underlying this regulatory convergence is the recognition by regulators that the risk management practices of banks are well ahead of those of other institutions. These other institutions – and especially insurance companies – are therefore being encouraged to see the risk management practices of banks as role models they should emulate.

Banks were the first set of institutions to be subject to a formal set of regulatory capital requirements. Banks' capital regulations were enshrined in 1988 Basel Accord which came into force in 1992 (Basel Committee (1988)). The Basel regime imposed two minimum standards of capital adequacy: an assets-to-capital multiple and a risk-based capital ratio of 8% of risk-weighted on-balance sheet assets plus off-balance sheet exposures, irrespective of the maturity or volatility of the values of the assets held. Two types of regulatory capital were permitted: tier 1 or core capital (equity and non-cumulative perpetual preferred shares less goodwill), and tier 2 or supplementary capital (subordinated debt with an original maturity in excess of 5 years and cumulative perpetual preferred shares).

However, the Basel regime soon revealed itself to be both naive and highly inadequate (see, e.g., Dowd (1997), Jackson et al (1997)). Three major weaknesses in particular stand out. The first is that the 8% ratio on which it was based was arbitrary: it was chosen to ensure that there no big jumps in most banks' regulatory capital requirements, and there was no attempt to justify in terms of it satisfying some desired target probability of bank insolvency. The second was that the regulations were littered with arbitrary provisions over such issues as netting arrangements, the so-called risk weights to be applied, and the like. These created considerable scope for regulatory arbitrage: banks could move into positions that were 'rewarded' by the regulations and away from positions that were 'penalised' by them, and in so doing reduce their regulatory capital charges. The third and possibly most serious weakness was in the building-block approach itself: by giving each asset a fixed 'risk weight', it implicitly contradicted the most basic principle of portfolio theory, namely, that the risk of a position is (special cases aside) not a function of the position itself, but a function of how the position relates to the rest of the portfolio of which it is a part. The whole notion of fixed 'risk weights' makes no sense if there is any diversification within a portfolio, i.e., it makes no sense unless one assumes that all risk factors are perfectly correlated. The adding-up approach therefore penalises globally diversified banks in comparison with specialists trading only in single asset classes. To add to which, there was also the difficulty that the risk weights were arbitrary and often made no market sense. So, for example, a position in UK or US government debt was assumed to be riskless, but this ignores the point that such positions are still exposed to market risk – the risk of falling asset prices – even if one assumes that any credit risk involved is negligible.

The original Basel Accord was also soon seen to be excessively rigid and dated. Its datedness was highlighted by the rapid rise of value-at-risk (VaR) models in the early 1990s and by widespread dissatisfaction with the fact that the regime did not allow banks to make use of these models in setting their regulatory capital requirements. However, after extensive discussions, an Amended Basel Accord – known as Basel 1 – was approved and came into effect in 1998.

This permitted banks a choice of two models for determining their trading book capital: they could use a version of the original 'standardised model' (or building-block model) and they could use the 'internal model' (or regulatory VaR) approach. This involved the use of a bank's own VaR model based on the VaR calibrated on the 99% confidence level and a 10-day holding or horizon period. The regulations imposed minimum standards on the sample sizes on which VaR estimates were to be made, on the types of back-test (or model validation tests) to be performed, and so on.

As an extra safeguard, Basel 1 set the capital requirement at the higher of the current VaR and a multiple of the average VaR over the previous 60 days. This multiplier was another arbitrary parameter set at between 3 and 4 at the discretion of the bank's regulatory supervisor based on his or her assessment of the model's backtesting performance, and was designed as a hedge against model and parameter risk, inaccurate assessments of credit risks, operational risks and unusual market moves. The models used for determining regulatory capital also had to be the ones that were actually used in the daily risk management of the bank.

The Amended Accord also allowed a third tier of sub-supplementary capital (short-term subordinated debt with an original maturity in excess of two years) to be allocated against the market risk of the trading book.

Banks had to first allocate tier 1 and tier 2 capital to meet credit risk capital requirements sufficient to cover 8% of risk-weighted assets. Then tier 3 capital was allocated to satisfy a second capital-assets ratio.[1] Thus, by 1998, banks had to satisfy three capital adequacy standards: a maximum assets-to-capital ratio of 20, a credit-risk capital charge of at least 8% of risk-weighted assets both on- and off-balance sheet, and a minimum market-risk capital charge to cover traded instruments in the trading book on and off balance sheet (see, e.g., Crouhy et al (1998)).

The acceptance of internal risk-based (IRB) models for determining capital requirements was a revolutionary departure for supervisors, but it was driven, in part, by the underlying complexity of the products in bank portfolios and the proprietary expertise required by banks in pricing and trading these products: the regulators recognised that capital requirements for banks' options and swap books could only be assessed sensibly by using the banks' own IRB models. However, the regulators were also aware of the importance of the need to supplement VaR modelling with stress tests, and regulations also stipulated some of the conditions that stress tests were to meet.

---

[1] The denominator of the new ratio is the sum of the risk-weighted assets and 12.5 times the market risk capital charge (where 12.5 is the reciprocal of the minimum capital ratio of 8%). The numerator of the second ratio is the sum of the bank's tier 1, tier 2 and tier 3 capital. Tier 3 capital could be used solely to meet the market risk capital charge, while tier 1 and 2 capital could also be used to satisfy the market risk capital charge once the credit risk allocation had been met in full. However, at least 50% of the bank's qualifying capital had to be tier 1 (with term subordinated debt not exceeding 50%) and the sum of tiers 2 and 3 capital allocated to market risk, not exceeding 250% of the tier 1 capital allocated to market risk (so that at least 28.57% of market risk capital had to be tier1).

Despite the theoretically superior structure of the internal model compared with the standardised model, some regulators still had doubts about the ability of banks to correctly model the principal (i.e., directional, spread, curve, and liquidity) risks contained in their portfolios (see, e.g., Kupiec and O'Brien (1995)). Credit rating agencies such as Standard & Poors (1996) also expressed concern that the amount of regulatory capital would fall as the high credit risk charge under the 1988 Accord for on-balance sheet holdings of bonds and equities would be replaced by a much lower specific risk capital charge under Basel 1. S&P argued that the market risks of trading operations are swamped by other more difficult to quantify risks such as operational risk (arising from systems failure and employee fraud), legal risk (arising from lawsuits from disgruntled clients), reputation risk, liquidity risk and operating leverage.

These and other concerns eventually led to a new capital adequacy regime known as Basel 2, which is to take effect in 2007 (see, e.g., Basel Committee on Banking Supervision (2003), Jackson (2001, 2002)). The stated objective of this new regime is to further enhance the security of the global banking system by introducing a three-pillar system of regulation that covers operational risk in addition to credit and market risk:

- Pillar 1 – sets minimum capital charges for credit, market and operational risks
- Pillar 2 – involves supervision by national financial regulators
- Pillar 3 – imposes market discipline via information disclosure

The regulatory capital held under pillar 1 must be sufficient to cover expected losses as well as unexpected losses. Again banks can choose to use an internal or standardised model for measuring risk, but all banks must have a robust risk identification structure in place by 2007 that categorises loans in terms of default bands.

As the regulatory capital required for each loan will depend on the probability of default (PD) of the borrower (determined by the bank) and the loss given default (LGD) (determined by the bank if it uses the advanced approach and set by the regulator if the bank uses the standardised approach). The expected loss to the bank is equal to the product of PD and LGP and the bank needs capital to cover at least its expected losses.

However, there has been substantial criticism of the requirement to hold regulatory capital to cover expected losses, particularly by US banks which argue that expected losses are a cost of doing business that should be incorporated into transaction prices and met from loan-loss provisions rather than covered in regulatory capital (see

*Global Risk Regulator*[2], September 2003). Some among the regulatory community have countered that it is hard to differentiate between expected and unexpected losses. It is also difficult to determine whether provisions are adequate to meet all expected losses, a necessary precondition for expected losses to be excluded from capital requirements.

Pillars 2 and 3 have also been subject to criticisms by the banking community. The banks are concerned that the discretion of national supervisors under Pillar 2 might be used to undermine the level playing field that banks world wide are supposed to play on. In respect of Pillar 3, the main concern is with the costs of providing the additional information required, measured against its usefulness to investors and regulators.

In addition, many other concerns have also been raised. For example, Basel 2 retains many of the weaknesses of Basel 1 – its inflexibility, and so on – and an abiding concern is that the regulations attempt to stipulate what best practice should be in an environment where best practice is evolving rapidly. The regulatory notion of best practice inevitably lags behind the market's perception of best practice, and regulatory requirements have often been criticised for hampering the development of 'true' best practice by imposing regulatory straightjackets on banks. A glaring example is the way in which regulatory requirements in effect lead to a situation where all banks that use the IRB approach to determine their regulatory capital are in effect being pushed to follow the same risk management strategies, i.e., to sell in a crisis, and such pressure is likely to exacerbate market instability in a crisis instead of reduce it.

**2.2 Life Assurance Companies**

In June 2004, the Financial Services Authority published the *Integrated Prudential Sourcebook for Insurers* (PSB)[3] which introduced a new set of risk-based capital requirements for with-profits life assurers that came into effect on 1 January 2005. The aim is to treat a life assurer's customers fairly through a combination of improved transparency, the holding of adequate capital for the firm's business mix and compliance with the Principles and Practices of Financial Management (PPFM) that the firm has disclosed. To achieve this aim and also to ensure that life assurers are treated in a way comparable to banks, the FSA has adopted the same three-pillar approach to regulation as Basel 2.

As noted earlier, pillar 1 of the three-pillar regime addresses regulatory capital and on this issue the PSB introduces a 'twin peaks' standard for a life assurer's with-profits business that results in provisioning and capital requirements being more responsive to the ways in which bonus payments are made to policyholders. The first peak is necessary to ensure compliance with the EU Life Assurance Directive, known as Solvency 1. This peak specifies the long-term insurance capital requirement (LTICR), defined as mathematical reserves plus a required minimum risk capital margin.

---

[2] http://www.globalriskregulator.com/
[3] This section draws heavily on this document, together with Consultation Paper 195 (Enhanced Capital Requirements and Individual Capital Assessments for Life Insurers) released in August 2003.

The mathematical reserves are the assets backing the life assurer's liabilities, calculated as the actuarial value of its contractual and guaranteed benefits. Net cash flows from current in-force business (benefits paid less premiums received) are forecast and discounted (using a discount rate that depends on the expected rate of return on fund assets) to give a net present value reserving requirement. Allowances, known as 'margins for adverse deviation', are made to account for potential forecast errors. For example, asset returns are reduced to allow for reinvestment risk and counterparty default risk, while the 'net premium rule' permits expected future premiums to be reduced by an amount that reflects the payment of future discretionary bonuses.

The risk capital margin (RCM) is the additional capital that a firm needs in order to maintain cover of its with-profit liabilities, given a sequence of specified stress events in market risk (equity, interest rate, and property price risk), credit risk (default by issuers of the firm's assets and non-payment by reinsurers), and persistency risk (the risk of a policyholder surrendering the policy early so that upfront marketing costs are not fully recouped). The specified stress events are:

- Equity:
    - For UK equities a fall of at least 10%, or if greater, the lower of:
        - A percentage fall in the market value of equities which would produce an earnings yield of the FTSE Actuaries All Share Index equal to 4/3rds of the long-term gilt yield; and
        - 25% less any percentage reduction between the current FTSE Actuaries All Share Index and its average over the last 90 days
    - Broadly equivalent test for overseas equities
- Interest rates: The more onerous of a fall or rise in yields on all fixed interest securities by a percentage point amount equal to 20% of the long term gilt yield (or comparable foreign government bond yield for foreign bonds)
- Real estate: A fall in real estate values of a minimum of 10% and a maximum of 20%; the required fall increases as the ratio of the current value of an appropriate real estate index to the average value of that index over the three preceding financial years increases.
- Credit risk:
    - Rated investment grade corporate bonds:
        - Increase in corporate bond yield spreads over equivalent risk free rates from spreads prevailing at valuation date. Increase by differential between current average bond yield spread and specified maximum bond yield spread. Maximum bond yield spreads of 90-210 basis points above risk free rates, according to credit grade of bond assets.
    - Rated, non-investment grade corporate bonds:
        - Increase in corporate bond yield spreads over equivalent risk free rates from spreads prevailing at valuation date. Increase by differential between current average bond yield spread and specified maximum bond yield spread. Maximum bond yield spreads of 525-900 basis points above risk free rates, according to credit grade of bond assets. For the lowest rated bonds, not in default, a fixed capital charge of 10% of the market value of that bond.

- Non-rated corporate bonds:
  - Where the firm assesses the credit quality to be equivalent to that of a rated bond, according to the rating and the method for corporate bonds. In other cases, a fixed capital charge of 10% of market value of bond.
- Commercial mortgages and other non-rated assets:
  - Where the firm assesses the credit quality to be equivalent to that of a rated bond, according to the rating and the method for corporate bonds. In other cases, a fixed capital charge of 10% of market value of the non-rated asset.
- Reinsurance concentration:
  - For material reinsurance arrangements,
    - where the reinsurer is rated, according to the credit rating of the reinsurer, and the method for corporate bonds;
    - where the reinsurer is not rated, a fixed capital charge of 10% of value of the reinsurance asset.
  - Intra-group reinsurance is excluded, where both insurer and reinsurer are regulated in a designated state.
- Assets in default, that are specifically provisioned in accordance with accounting practice:
  - No credit stress required
- Persistency:
  - Termination rates in each year of projection of 50% of the termination rates assumed in realistic liabilities.

The RCM is the sum of the net losses for each of the above scenarios (i.e., in the cases where the assets fall by more than the fall in with-profit liabilities), and is required to be at least 4%.

There is also a resilience capital requirement applied to the mathematical reserves. This test requires additional resilience capital to be set aside if stressed market conditions indicate that asset values will fall by more than the reduction in mathematical reserves. The modification is that the additional capital can come directly from shareholders' capital and need no longer be included in the mathematical reserves which are held in the life fund itself. In addition, the LTICR is calculated with reference to mathematical reserves net of resilience capital.

The second peak is based on a realistic calculation of with-profits liabilities by life assurers. The PSB permits firms to carry out such calculations using one of two methods: the asset share approach and the prospective or bonus reserve approach. For the purposes of valuing contracts with guarantees and embedded options, the PSB permits stochastic valuation and option pricing models. If the second peak is higher than the first peak, additional capital (the with-profits insurance capital component, or WPICC) will be needed to cover expected discretionary bonus payments (such as annual increases in reversionary bonuses and the terminal bonus). The WPICC makes an allowance for adverse experience: the future values of realistic assets and liabilities might be respectively less or more than expected as a result of a firm's exposure to market, credit and persistency risks.

The FSA estimates that the required level of pillar 1 capital (the capital resources requirement, or CRR) is the same as would attract a Standard and Poors BBB rating. A BBB rated insurer 'has *good* financial security characteristics, but is more likely to be affected by adverse business conditions than are higher rated insurers'. The FSA suggests that this equates to a 99.5% confidence level that the firm concerned will survive for a one-year period.

The second or supervisory pillar is handled in the PSB by a framework of individual capital adequacy standards (ICAS). Each firm assesses the level of capital suitable for its own risk profile (the individual capital assessment, ICA) and this is then compared with the minimum capital requirements for with-profits business (established by the twin peaks standard) and the insurer's other life business.

The PSB provides general guidance on the risks to capital that life assurers should consider in relation to their individual capital needs. It also provides guidance on how the risks might be assessed by means of capital stress tests, scenario analyses, or other models (such as economic capital models).

The FSA also offers individual capital guidance (ICG) in the light of a life assurer's individual capital assessments. To do this the following information needs to be submitted to the FSA:

| Item | Coverage |
|---|---|
| Summary | A summary of the financial position of the firm at the time the report is constructed and the risks to which the firm is subject to. |
| Individual capital assessment (ICA) | The firm's proposed ICA, expressed as a proportion of its 'pillar 1' capital resources requirement (CRR). |
| Background | Relevant historical development of the firm and any conclusions that can be drawn from that development which may have implications for the future of the firm. |
| Current business | The current business profile of the firm. |
| The future | The environment in which the firm expects to operate, and its projected business plans, projected financial position and future sources of capital. |
| Capital analysis | A detailed review of the capital adequacy of the firm. This analysis could include a commentary and opinion on the applicability of the CRR to the firm's own capital position and its appropriateness compared to its own capital assessment. It could involve an analysis of current capital levels and movements in solvency during the past years, future capital requirements and general outlook. |

| Risk assessment | An identification of the major risks faced in each of the following categories: credit risk; market risk; insurance risk; operational risk and liquidity risk; and the extent to which the firm holds capital in response to each risk. |
|---|---|
| Stress and scenario tests | The quantitative results of stress and scenario tests carried out by the firm and the confidence level and key assumptions behind those analyses. |
| Other risks | Identification of any risks, for example systems and controls weaknesses, which in the firm's opinion are not adequately captured by the CRR. The firm's assessment of how it is responding to those risks, and if through holding capital, the amount. |
| Capital models | If a more sophisticated modelling approach is used by the firm, we would expect a statement of the confidence level and other parameters that have been used in the model. |
| Source: FSA CP195, p 48, 2003 | |

Given this information, the FSA will confirm either that the firm's capital assessment is adequate or that a higher level of capital is required in the light of the FSA's judgement that the firm's business risks are greater than the firm has itself assessed.

The FSA wishes to be confident that if the projected adverse financial situations materialise, then firms will still be able to pay their liabilities in full when they fall due. This requires that assets are valued at their liquidation value under the relevant scenarios and that liabilities are given a realistic value for their due date. The overarching aim is to ensure that a firm's customers are 'treated fairly'. This means that a firm must have sufficient resources to ensure that its customers' 'reasonable expectations' concerning terminal bonuses are fulfilled.

The PSB also takes account of the EU's 'Solvency 1' Life Directive in the following ways. The minimum capital requirement for life assurers is set at Euro 3 million and this will be updated in line with EU consumer price inflation. The capital resource requirement (or required margin of solvency) must be met at all times rather than just at the date of the last balance sheet. It can be met with ordinary shares without limit and with cumulative preference shares, subordinated debt and unpaid share capital up to specified limits (and in the last case with approval):

| Tier of Capital | Limit |
|---|---|
| *Tier 1* | |
| Core tier 1 | Unlimited but at least 50% of total tier 1 |
| Ordinary shares | |
| Reserves | |
| Non-ordinary shares | |
| Innovative tier 1 | 15% of total tier 1 |
| Capital instruments | |
| Innovative instruments | |
| Implicit items | Waiver required |
| *Tier 2* | 100% of total tier 1 |
| Upper tier 2 | |
| Perpetual cumulative preference shares | |
| Perpetual subordinated debt | |
| Lower tier 2 | 25% of total capital resources |
| Long term subordinated debt | |
| *Other capital* | Waiver required |
| Unpaid share capital | |
| Source: FSA CP195, p 62 | |
| Note: Characteristics of innovative instruments (FSA CP195, p 63, 2003): Treated as a liability in financial statements; coupon payments may be deferred with any deferred coupons payable only in shares; no specified redemption date but terms may include an issuer call which may coincide with an increase in the coupon; normally ranks pari passu with preference shares; loss absorbency usually achieved through conversion into shares at a predetermined trigger event. | |

Currently future profits can be used to offset the capital requirement. Implicit items for future profits are restricted to 2/3rds of the firm's LTICR (or to the level of the LTICR minus EUR 3 million, if less). By 2007 implicit items for future profits must be restricted to 25% of the lesser of the LTICR and its total (eligible) capital resources; and from 31 December 2009 they will no longer be allowed. Firms will be required to submit an actuarial report substantiating the emergence of anticipated profits in future periods.

The PSB also changes the way in which capital resources are reported. The traditional approach measures the total of admissible assets less foreseeable liabilities. The new approach lists the components of capital. Both calculations give the same result as shown in the following table (drawn from FSA CP195: Table 2.2.10 G, Annex 6, 2003):

| **Liabilities** | | **Assets** | |
|---|---|---|---|
| Borrowing | 100 | Admissible assets | 350 |
| Ordinary shares | 200 | Intangible assets | 100 |
| Reserves | 100 | Other inadmissible assets | 100 |
| Perpetual subordinated debt | 150 | | |
| Total | 550 | Total | 550 |

| Traditional calculation of capital resources: Eligible assets less foreseeable liabilities | |
|---|---|
| Total assets | 550 |
| Less Intangible assets | 100 |
| Less Inadmissible assets | 100 |
| Less Liabilities (borrowing) | 100 |
| Capital resources | 250 |

| New calculation of capital resources: Components of Capital | |
|---|---|
| Ordinary shares | 200 |
| Reserves | 100 |
| Perpetual subordinated debt | 150 |
| Less Intangible assets | 100 |
| Less Inadmissible assets | 100 |
| Capital resources | 250 |

This new approach is the same as that used by banks, building societies and investment firms.

For its part, the EU is introducing a 'Solvency 2' Life Directive in 2007 which will bring the regulation of life assurers even closer to the three-pillar Basel 2 framework of capital charges, supervisory review and information disclosure.[4]

So although there are legitimate criticisms of the regulatory regimes that have been established for both banks and life assurers, what is nevertheless clear is the extremely detailed exercises these institutions need to perform to determine the regulatory capital they need to (hopefully) remain solvent with a high degree of probability.

---

[4] There are also important accounting issues to consider. In 2005, the International Accounting Standards Board introduced 'fair value' international accounting standards. However, insurers are required to use the fair value standard for assets from 2005 but the fair value standard for liabilities (which replaces the book value measure) is to be used from 2007. The FSA believes that fair-value accounting is a necessary concomitant of risk-based solvency regulations: in order to ensure that firms match their capital more accurately to the risks they face, they need to measure both their assets and liabilities in a fair and transparent manner, in contrast with the opaque valuation methods that were commonly used. Needless to say, there has been intensive debate on the impact of 'fair value' accounting, and the main argument against is the concern that it might lead to excessive spurious earnings volatility.

## 3  The Financial Regulation of Pension Funds: Why is it different?

The financial regulation of UK pension funds differs from that of banks life assurers in a number of key respects. First, they are not regulated by the FSA at all: they are regulated by the Pensions Regulator (TPR) which is not even based in the same building as the FSA but in a different city. Second, pension funds do not have any formal capital requirements: instead they operate on a prudent person principle which gives pension fund managers much greater discretion than their counterparts in other financial institutions. This raises the question of why the regulatory system does not treat pension funds in the same way as other financial institutions, including institutions that also make long-term investments, such as life assurers.

Defined benefit pension schemes share many of the characteristics of the with-profit policies sold by life assurers. Both aim to deliver a pre-determined benefit (a fixed minimum return in the one case, a fixed proportion of final salary in the other), despite investing in assets whose returns can be highly volatile. It is therefore curious that whilst the financial regulation of insurance companies is moving closer to that of banks, the financial regulation of pension funds remains very different to either.

The third key difference between banks, life assurers and pension funds is that the first two are subject to a solvency standard whereas the latter is subject to a funding standard. A solvency standard ensures that assets exceed liabilities. A funding standard involves setting a smooth path for contributions that is meant to enable the fund to pay the promised benefits over the long run.[5] A funding standard is much weaker than a solvency standard. For example, a pension fund can be fully funded but still be unable to pay its liabilities in full if the sponsor becomes insolvent. This means that a fully funded scheme might still depend on future sponsor contributions to make good any deficit.

The first attempt at the financial regulation of pension funds was the Minimum Funding Requirement (MFR) introduced by the 1995 Pensions Act. The MFR (which came into effect on 6 April 1997) established a minimum level of funding for a defined benefit (DB) pension scheme (or for a defined contribution pension scheme which also provides salary-related benefits) and an associated schedule of contributions necessary to meet this minimum level of funding. The pension scheme's trustees were responsible for ensuring that this schedule was delivered. The MFR could be satisfied either by the minimum level of funding being met immediately or by having a schedule of contributions in place that would meet the minimum funding level within a specified time limit (initially a maximum of 5 years, subsequently extended to 10 years).

---

[5] Underlying regulatory thinking here is the 'prudent person' principle. To quote paragraph (31) of the Preamble to the new European Union Pension Fund Directive (2003):
"Institutions are very long-term investors. Redemption of the assets held by these institutions cannot, in general, be made for any purpose other than providing retirement benefits. Furthermore, in order to protect adequately the rights of members and beneficiaries, institutions should be able to opt for an asset allocation that suits the precise nature and duration of their liabilities. These aspects call for efficient supervision and an approach towards investment rules allowing institutions sufficient flexibility to decide on the most secure and efficient investment policy and obliging them to act prudently. Compliance with the 'prudent person rule' therefore requires an investment policy geared to the membership structure of the individual institution for occupational retirement provision."

A pension scheme was defined to have a 'deficiency' when it has insufficient assets to meet its liabilities. The schedule of contributions needed to make good any deficiency must be agreed between the trustees and sponsor. A 'serious deficiency' occurs when the assets are valued at less than 90% of the value of the liabilities. To reduce such a deficiency, the assets had to be increased to at least 90% of the liabilities, valued on the basis set out under the MFR rules within one year (later extended to three years). This outcome could be achieved either through a cash payment to the fund by the sponsor or by the sponsor giving a financial guarantee to bring the scheme's assets up to at least 90% of the liabilities in the event that the sponsor becomes insolvent and contributions to the fund must continue to be paid. If neither of these solutions was feasible, the trustees had to inform the Occupational Pensions Regulatory Authority (OPRA)[6] within 14 days and scheme members within one month.

If the deficiency was less serious and assets were worth between 90% and 100% of liabilities, then the assets had to be increased to 100% of the liabilities by the end of the period covered by the schedule of contributions. Contributions might be increased to achieve this outcome, and any such increased contributions could be spread evenly throughout the period covered by the schedule. It was also permissible for larger contributions to be paid early on in the period (this is called 'frontloading'), but the 'backloading' of contributions towards the end of the period was not permitted.

Following each MFR valuation, the trustees had to establish a schedule of contributions within twelve weeks. Each schedule covered a five-year period and might need to be revised during this period to ensure that the MFR continued to be met. The schedule showed the contribution rates and due dates for all the contributions to be paid:

- by (or on behalf of) all active members (excluding additional voluntary contributions)
- by (or on behalf of) each sponsoring employer taking part in the scheme
- by the sponsoring employer to rectify a serious shortfall in funding.

Even with this schedule of contributions, it was not necessarily the case that the whole of a scheme's liabilities could be met in full if the scheme were to be wound up immediately. The MFR did not guarantee absolute security for pensions because (unlike Basel 2 and Solvency 2), the MFR was a funding standard and not a solvency standard.[7] As the Chairman of the Pensions Board of the Faculty and Institute of Actuaries (FIA), Mike Pomery, stated at the 2000 NAPF annual conference, the MFR gave scheme members only a 'reasonable expectation' that they would get their full pension, not 'absolute security'.

In any case, the FIA estimated that full funding for UK pension funds (i.e., the full cost of a buy-out with insurance companies) would cost an additional £100bn on top of assets valued at £830bn in 2000 (Faculty and Institute of Actuaries (2000)). There were a number of reasons for this:

---

[6] The predecessor to TPR between 1997 and 2005.
[7] The Goode Report (1993), which led to the 1995 Pensions Act that established the MFR, recommended that pension funds should be subject to a solvency standard, but this recommendation was watered down into a much weaker funding standard by the time the Act was passed.

- the claims of retired members were met first
- the insurance companies that provided both immediate and deferred pension annuities for members when a sponsoring company was wound up were likely to use lighter mortality assumptions than allowed for in the MFR regulations and hence offered lower annuities for a given purchase price
- falling long-term interest rates since 1990 which raised the present value of scheme liabilities; even though the assets held by DB schemes, mainly equities, had traditionally delivered very high returns, they still failed to keep up with the growth in scheme liabilities since the introduction of the MFR in 1997.
- liabilities were valued using the current unit method with LPI revaluation, and so did not take into account future earnings growth.

As many as one in six pension funds in 2000 were either at, or below, the MFR borderline of 90% funding. The weakness of the MFR standard was exposed in 2000 by the case of Blagden, a chemicals company whose pension fund fully satisfied the MFR, but which went into insolvency with funds sufficient only to meet two-thirds of its obligations to active members.

The resulting public debate led to a Treasury-sponsored review of institutional investment chaired by Paul Myners, chief executive of Gartmore. The Myners Report was published in March 2001 and its recommendations were immediately accepted in full by the Government (Myners (2001)). Myners called for a new approach to institutional investment, identified a series of current distortions to effective decision-making, and suggested ways of tackling these distortions.

One of the key features of the report was its proposal to replace the MFR with a long-term scheme-specific funding standard in the context of a strong regime of transparency and disclosure. The report also proposed a set of additional measures to strengthen protection:

- a recovery plan for returning schemes to full funding
- a statutory duty of care on the scheme actuary
- stricter conditions on the voluntary wind-up of a scheme where the employer remains solvent (e.g., the liabilities would have to include the actual cost of winding up the scheme and the actual cost of buying annuities to secure pensions in payment), and
- an extension of the fraud compensation scheme: the level of compensation for fraud would be increased to cover not simply the MFR liabilities as at present, but the full cost of securing members' accrued benefits (or the amount of the loss from fraud, whichever is the lesser).

As the Government noted, 'These proposals will provide protection for members of all defined benefit schemes and will encourage an intelligent and thought-through approach to planning investment and contributions policy. They do not distort investment as the MFR does, because they do not involve the valuation of liabilities using statutory reference assets which create artificial incentives for schemes to invest in those assets. Employers that wish to go on offering defined benefit schemes will find it easier to do so under these proposals. At the same time, the proposals will make it more difficult for those that wish to walk away from the pension promises that they have made'.

On 11 June 2003, the Government announced that any solvent company which wound up its DB pension scheme had to do so, not on an MFR basis, but on a full buy-out basis with a life assurer (i.e., the fund had to have sufficient assets to buy immediate annuities for the scheme's pensioners and deferred annuities for active and deferred members).

Following this, the 2004 Pensions Act introduced the requirement for a scheme-specific funding standard[8] to replace the MFR. This requires scheme trustees to:
- prepare a Statement of Funding Principles[9] specific to the circumstances of each scheme, setting out how the Statutory Funding Objective (SFO)[10] will be met
- obtain periodic actuarial valuations and actuarial reports
- prepare a schedule of contributions
- implement a recovery plan where the SFO is not met.

Trustees are also required to prepare a transparency statement that reports:

- the current value of its assets and in what asset classes they are invested
- the assumptions used to determine its liabilities
- planned future contributions
- its planned asset allocation for the following year or years
- the assumed returns and assumed volatilities of those returns for each asset class sufficient to meet the liabilities
- a justification by the trustees of the reasonableness of both their asset allocation and the investment returns assumed in the light of the circumstances of the fund and of the sponsor, and
- an explanation of the implications of the volatility of the investment values for possible underfunding, and a justification by trustees of why this level of volatility is judged to be acceptable.

---

[8] Occupational Pension Schemes (Scheme Funding) Regulations 2005 (S.I. 2005/3377).

[9] The Statement of Funding Principles sets out trustee policy for meeting the Statutory Funding Objective and should include:
- funding objectives and the trustees' policy for meeting it
- the scheme's investment policy
- whether the Regulator has given any direction in relation to the scheme
- the calculation basis for measuring assets and technical provisions
- how often actuarial valuations will be obtained
- how cash equivalent transfer values will be calculated.

[10] The Statutory Funding Objective states that the scheme must have sufficient and appropriate assets to cover its technical provisions. The technical provisions are an estimate, made using actuarial principles, of the assets needed at any particular time to make provisions for the benefits that have already accrued under the scheme, including pensions in payment, benefits payable to the survivors of former members and those benefits accrued by other members which will be payable in the future. The technical provisions are calculated using an accrual benefits funding method and assumptions all chosen by the trustees, after taking the actuary's advice and obtaining the employer's agreement (Pensions Act 2004).

The scheme-specific funding standard must reflect the specific liabilities of the scheme. This suggests that the pension fund should be looking to invest in assets that match as closely as possible the liabilities of the scheme in terms of key features of the liabilities, such as the way that they change over time in response to earnings growth, changing interest rates and demographic factors, such as the maturity structure of the liabilities of the scheme.

**4 Lessons from Other Government-Sponsored Insurance Schemes**

**4.1 Financial Services Compensation Scheme**

The Financial Services Compensation Scheme (FSCS) is a UK Government-sponsored insurance scheme covering companies authorised by the FSA. It came into operation on 1 December 2001.[11] The FSCS is independent of the FSA and provides compensation to consumers if an authorised company becomes insolvent and is not able to pay its liabilities.

The FSCS provides for three kinds of compensation, with different rules and limits:

- Deposit Claims include deposits with banks, building societies and credit unions.
- Insurance Claims include:
    - compulsory insurance (such as third party motor insurance)
    - non-compulsory insurance (such as home insurance)
    - long-term insurance (such as pension plans and life assurance).
- Investment Claims include claims relating to bad investment advice or poor investment management, or where a firm has gone out of business and cannot return your investments or money.

The compensation limits for the FSCS are: deposits £31,700 (100% of £2,000 and 90% of £33,000); long–term insurance at least 90% of the value of the policyholder's guaranteed fund at the date of default; general insurance, compulsory, 100% of valid claim/unexpired premiums, non compulsory, 100% of the first £2,000 of valid claim/unexpired premiums and 90% of the remainder of the claim; investments £48,000 (100% of £30,000 and 90% of next £20,000).

The FSCS is funded by levies on the industry on a pay-as-you-go basis. There are two types of levy: a management expenses levy and a compensation costs levy. The former is determined annually in advance and covers 'base costs' (payable by all firms) and the 'specific costs' associated with paying compensation which depends on the number of claims and types of default.

---

[11] The FSCS replaced eight existing schemes each of which provided compensation if a firm collapsed owing money to depositors, policyholders or investors: the Deposit Protection Scheme, the Building Society Investor Protection Scheme, the Policyholders Protection Scheme, the Friendly Societies Protection Scheme, the Investors Compensation Scheme, the Section 43 Scheme (which covers business transacted with listed money-market institutions), the Personal Investment Authority indemnity scheme, and the arrangement between the Association of British Insurers and the Investor Compensation Scheme Ltd for paying compensation to widows, widowers and dependants of deceased persons.

The compensation costs levy is also determined in advance on the basis of 'anticipated compensation costs for defaults expected to be declared in the 12 month period following the date of the levy'. These include 'the costs incurred in paying compensation, securing continuity of long-term insurance and safeguarding eligible claimants when insurers are in financial difficulties' (Financial Services Authority (2001)).

The FSCS is supported by the regulatory frameworks facing the institutions it covers. These determine the minimum regulatory capital needed to cover specified losses. In principle, these enable the probability of loss and expected loss to be quantified and hence enable the insurance premium for the FSCS to be set on the basis of standard insurance principles. The FSCS is therefore likely to be an effective insurance scheme whose solvency is assured by the adequacy of its premium income.

**4.2 The Pension Benefit Guaranty Corporation (PBGC)**

We now turn to the United States and, in particular, to the Pension Benefit Guaranty Company. This is particularly relevant to us here as the PBGC is the model on which the Pension Protection Fund is based.[12]

The PBGC was created by the Employee Retirement Income Security Act of 1974 to encourage continuation and maintenance of DB pension plans in the US, provide timely and uninterrupted payment of pension benefits, and keep pension insurance premiums at a minimum. The PBGC covers 44 million workers in 30,000 DB plans. It pays monthly retirement benefits to 683,000 retirees in 3,600 pension plans that have ended and is responsible for the current and future pensions of about 1.3 million people. For plans ended in 2006, workers who retire at age 65 can receive up to $47,659 a year.

The PBGC's premium revenue was $1.5 billion in 2005. All single-employer pension plans pay a basic flat-rate premium of $19 per participant per year. Underfunded pension plans pay an additional variable-rate charge of $9 per $l,000 (i.e., 0.9%) of unfunded vested benefits. The premium for smaller multiemployer program is $2.60 per participant per year. The PBGC paid nearly $3.7 billion in benefits in 2005. The sense among experts in the field is that these premiums are worryingly low in comparison with the potential payouts expected of the PBGC: in 2005, the PBGC had a deficit of liabilities over assets of $23 billion.

An interesting aspect of the PBGC is how the premium it charges has increased since 1974:

- 1974: flat-rate of $1 per participant
- 1978: flat-rate of $2.60 per participant
- 1986: flat-rate of $8.50 per participant
- 1988:

---
[12] See Pension Benefit Guaranty Corporation website http://www.pbgc.gov/.

- basic premium raised to $16
- additional variable-rate premium was imposed on underfunded plans of $6 per $1,000 of unfunded vested benefits up to a maximum of $34 per participant
- 1991:
  - basic premium raised to $19
  - additional variable-rate premium raised to $9 per $1,000 of unfunded vested benefits up to a maximum of $53 per participant
- 1994:
  - basic premium stays at $19
  - variable-rate premiums increased for plans that pose greatest risk by phasing out maximum limit on premiums for underfunded plans
- 1996: maximum variable-rate premium completely eliminated
- 2006: basic premium raised to $30

Although the premium is exposure-related (i.e., it is related to the level of the claim in the event of insolvency), the premium is not explicitly risk-related (i.e., so it is not higher for sponsors more likely to become insolvent). The premium is also not related to the probability of a claim being made. This means that, contrary to standard insurance principles, financially weak sponsors with underfunded schemes are not charged the full risk-adjusted premium. This is of course a major weakness.

There are three ways in which a pension plan can be taken over by the PBGC.

The first is 'distress termination'. A company in financial distress might voluntarily terminate a pension plan if: a petition has been filed seeking reorganization in bankruptcy, it has been demonstrated that the sponsor or affiliate cannot continue in business unless the plan is terminated, or it has been demonstrated that costs of providing pension coverage have become unreasonably burdensome solely as result of the decline in number of employees covered by the plan.

The second is 'involuntary termination'. The PBGC may terminate a pension plan if: the plan has not met minimum funding requirements; the plan cannot pay current benefits when due; a lump sum payment has been made to a participant who is a substantial owner of the sponsoring company; or the loss to the PBGC is expected to increase unreasonably if the plan is not terminated.

The third is 'standard termination'. A plan may terminate if the plan assets are insufficient to satisfy all plan benefits e.g., through the purchase of annuities with an insurer. There were 166,522 standard terminations between 1974 and 2005.

Perhaps not surprisingly, the PBGC has experienced a number of cases in which companies have deliberately underfunded their pension plan in advance of their own bankruptcy. It has sought to protect itself against such behaviour through a number of defences.

One of these defences is an Early Warning Program. The PBGC monitors certain companies which are financially distressed or have underfunded DB plans to try to prevent losses before they occur, rather than waiting to pick up the pieces afterwards. The PBGC will then contact a company if: (1) the company has a below-investment-grade bond rating and sponsors a pension plan with a current liability of over $25m or (2) the company (regardless of its bond rating) sponsors a pension plan that has a current liability over $25m and that plan has an unfunded current liability over $5m. It is particularly concerned about transactions that substantially weaken the financial support for a pension plan such as the breakup of a controlled group, the transfer of significantly underfunded pension liabilities in connection with the sale of a business, or a leveraged buyout.

Once the PBGC has identified a potential transaction that could jeopardize the pension insurance program, it meets with corporate representatives to negotiate additional contributions or security. It will work with the company to find a settlement appropriate to financial feasibility of company. However, in the event of the company becoming insolvent and the PBGC taking over the plan liabilities, the PBGC can claim up to 30% of the company's net worth to cover a deficiency in the plan.

It is, of course, open to question whether such measures really give the PBGC sufficient protection to survive: Figure 1 shows how rapidly claims against the PBGC have built up over the last few years. However, there is no denying that the PBGC does seek to identify potential problems in advance and that it has developed specialized tools – including specialised technology, databases, financial expertise, co-ordination with other regulatory and governmental bodies, etc. – to help it operate.

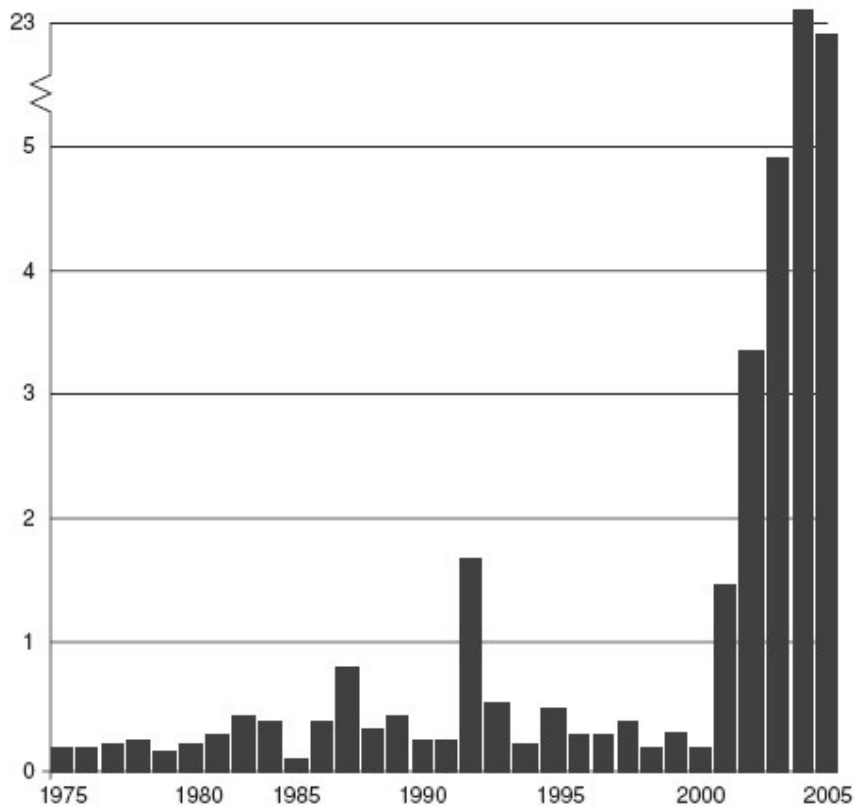

**Fig 1 Claims on the PBGC 1975 – 2005 (single employer plans) ($bn)**

So we have mixed lessons for the PPF in respect of the FSCS and PBGC. The former covers companies subject to a rigorous solvency standard and hence has a good chance of being able to meet claims if any of these companies does become insolvent. The latter currently has liabilities way in excess of its assets. We now turn to examine the risks faced by the PPF itself.

**5 The Financial Risks facing the PPF**

As we mentioned above, pension schemes in the UK have traditionally operated on a prudent person principle and made promises rather than guarantees. This is the principal reason why they have not faced formal capital requirements, in contrast with, say, life assurers which do offer contractual guarantees. However, the establishment of the the PPF is intended to mark a shift from promises to guarantees. Speaking at the Labour Party annual conference on 1 October 2003, the Chancellor of the Exchequer Gordon Brown said the Government would 'legislate for a new statutory pension protection fund. In future every worker contributing to a pension will have their pension protected and be *guaranteed* their pension rights'. If we take Mr. Brown at his word, this means that a *promise* made by a scheme sponsor is ultimately *guaranteed* by the PPF.

This will result in the PPF being subject to three key risks which we now examine.

**5.1 Moral hazard**

The first of these is moral hazard. This is one of the classic risks facing all insurance providers: people become more careless once they are insured and also have an incentive to play 'games' against the insurer. The PPF provides scheme sponsors with an incentive to underfund their schemes and invest in assets with higher expected returns and risks (Utgoff (1993)). This is because the value of the PPF guarantee is greatest for 'those schemes where the sponsor is financially weak, the pension scheme is poorly funded, the equity exposure is high and contributions are low' (McCarthy and Neuberger (2005)). If the assets perform well, the deficit will be reduced, but if the assets perform badly and the scheme becomes insolvent, the PPF will take over the pension liabilities. The PPF can respond by increasing the risk-based premium for an underfunded scheme, but this might not solve the problem and might actually make matters worse for a sponsoring company which is already in financial difficulty. For an employer near to insolvency, there is a pronounced trade-off between pensions and jobs.[13]

Furthermore, the very existence of the PPF provides an incentive for financially weak companies to increase pension benefits rather than wage increases: after all, the latter have to be paid immediately, while the former might eventually be paid by the PPF. Other 'games' the PPF should be wary of are the early retirement of senior directors of a company taking substantial pension benefits with them, the sale of a subsidiary with an underfunded pension scheme to a financially weak buyer, and pressure on the scheme actuary to change the actuarial assumptions in a way that lowers the reported

---

[13] Certainly for risk-based premiums to have the desired effect of increasing the funding level, the premium must be greater than the cost to the sponsor of borrowing funds to reduce the deficit.

deficit and reduces employer contributions (see also Gebhardtsbauer and Turner (2004)).

And, indeed, the Government should protect itself against moral hazard on the part of the PPF itself. The PPF too has an incentive to take increased asset risk, since it can rely on future premiums and an implicit (albeit denied) Government underwrite. Whatever the Government might say about the PPF fending for itself, the fact is that the Government would be reluctant to allow the PPF to go bust, and this reluctance exposes the Government to the danger of moral hazard by the PPF.

**5.2 Adverse selection: Bad drives out good**

Another classic risk that the PPF is exposed to is adverse selection: only those most likely to claim take out insurance. As with all forms of insurance, strong schemes subsidise weak schemes. The PPF is exposed to this risk because the levies charged by the PPF provide a strong incentive for financially strong employers to close down their defined benefit schemes, leaving only the schemes of financially weak sponsors participating in the PPF. Although participation is mandatory if an employer has a DB scheme, there is no requirement for an employer to operate a DB scheme in the first place. For the last ten years, firms have been switching away from DB schemes towards DC schemes, although on 11 June 2003, the Government announced that solvent companies could not walk away from their DB obligations accrued before this date unless they were fully bought out by an insurance company by means of current and deferred annuities. It is also possible that some strong firms will seek to avoid the PPF level, and there is anecdotal evidence that some companies are considering taking their businesses offshore for just this reason.

**5.3 Systemic risk**

Insurance works best where the risks covered by the insurer are specific or idiosyncratic risks, i.e., risks that are uncorrelated across claimants. This is because specific risks can be pooled and the insurance book is diversified. Insurance works less well if the risks assumed are systemic: in such cases there is little benefit from diversification.

Unfortunately, there are good reasons to think that some of the risks faced by the PPF are indeed systematic. For example, McCarthy and Neuberger (2005) make the point that the PPF faces systemic risk because insolvencies are cyclical. Claims arise when firms become insolvent and the claim size depends on the level of under-funding. Since pension funds have a heavy equity exposure, under-funding is therefore worst after sharp falls in stock markets and this is just when corporate insolvencies are likely to peak.

McCarthy and Neuberger support this argument using the results of an illustrative simulation model. They assume for the sake of argument that pension funds invest two-thirds of their assets in equities, have a 10-year deficit amortisation period, guarantee 100% of their liabilities and set premiums equal to the corresponding average annual breakeven claim rate of 0.3% of liabilities. Given these assumptions, their model suggests that, over 30 years, it is *likely* that there will be one year in which the claim rate is 1.2% and that it is plausibly *possible* that claims could equal 10% of liabilities. In other words, even if one accepts that claims will be low on average, the PPF is likely to experience years when the claims will be very high when prolonged weakness in equity markets coincides with widespread corporate insolvencies.

They go on to argue that it will be hard for the PPF to build reserves to cover claims of this size and therefore the PPF will need to raise premiums sharply after a prolonged market downturn, i.e., at the very time when companies will be financially stretched. In such a situation, there is a very real danger that the PPF could become insolvent.[14]

**6 Dealing with these Risks**

Naturally, the PPF has a strong incentive to design the insurance it provides to protect itself against these problems as best it can. Were the PPF a standard insurer, it might consider any of the following possible defences (drawing lessons where possible from the earlier sections of the paper):

- the PPF could be permitted to convert its claim against the sponsoring company from a debt claim to an equity claim
- having a maximum payout following a successful claim (i.e., co-insurance)
- permitting risk-based insurance premiums linked to the level of plan underfunding
- having a funding standard for schemes that will limit risk taking by the sponsoring company
- close supervision and threatening the public exposure of companies that are underfunding their schemes.

We will examine whether these defences are likely to be successful in the case of the PPF.

**6.1 An equity claim against the sponsoring company**

This defence will not, in general, work for the PPF as the sponsoring company which puts its pension scheme into the PPF will itself be insolvent.[15]

**6.2 A maximum payout**

Limiting the payout to a maximum proportion of the liabilities is generally a good way of reducing moral hazard. Unfortunately, this defence cannot be applied by the PPF because the 2004 Pensions Act specifies that the size of the payout should be independent of the assets in the fund.

---

[14] There is in fact a fourth risk that the PPF is also subjected to: political risk. For example, financially weak companies could exert pressure on the politicians in the constituencies where they are located to press for a reduction in the premiums that they face. There is also a risk that the Government limits contributions into the pension scheme in good economic times in order to limit its tax loss and in doing so limits the surplus that provides a cushion against later falls in equity markets. In fact, the 1986 Finance Act did precisely this and limited pension scheme surpluses to 5% of liabilities.

[15] However, in 2005, the Pension Protection Fund (PPF) did buy 10% of insurance broker Heath Lambert in return for bailing out its pension scheme.

However, other restrictions might help in certain circumstances. For example, the PPF might refuse to cover pension benefit withdrawals (especially by senior directors) or benefit increases made in a specified period (such as three years) prior to a scheme's insolvency. Similarly, the PPF might refuse to permit the sale of a subsidiary with an underfunded pension scheme to a financially weak buyer without a guarantee from the parent company.[16]

**6.3 Risk-based premiums linked to the level of underfunding**

Risk-based premiums are often suggested as a good way of dealing with moral hazard, but there are two problems with risk-based premiums in this case.

The first deals with assessing the correct level of default risk and hence premiums. Bodie (1996) shows that the default risk facing the PPF depends on the financial strength of the sponsoring company, the level of underfunding of the scheme, and the extent of mismatch between the scheme assets and liabilities. The problem is that the last two factors depend on the sponsor's contribution policy and the scheme's investment strategy, respectively. So, for example, an underfunded scheme might follow an aggressive equity-based investment strategy, hoping to rely on the equity risk premium to compensate for its inadequate contributions.

More light can be shed on these issues by thinking of the insurance provided by the PBGC or PPF as a put option on the scheme's assets (see, e.g., Sharpe (1976), Treynor (1977), Langetieg et al (1982), Marcus (1987) and Lewis and Pennacchi (1999a,b)). Vanderhei (1990) found that the insurance premiums charged by the PBGC significantly underestimated the true level of risk assumed by the PBGC, despite the fact that the basic premium had increased 30-fold since the PBGC was established. More specifically, Vanderhei estimated the size of the insurance premiums the PBGC needed to charge to cover its costs. As mentioned above, in order to be actuarially fair, this must equal the expected loss to the PBGC which, in turn, is equal to the product of the probability of default (PD) of a pension scheme and the loss given default (LGD). Using data supplied by the PBGC, Vanderhei calculated the breakeven insurance premium for the PBGC using its own formula of a fixed premium per member and a variable premium per $1,000 of underfunding. He found that the PBGC was undercharging on average (by a factor of 4.5), but also imposed significant cross-subsidies from strong to weak firms, thus worsening the problem of adverse selection.

The second problem is whether risk-adjusted premiums will have the desired effect of reducing risk-taking. Both McCarthy and Neuberger (2005) and Gebhardtsbauer and Turner (2204) give persuasive arguments to doubt that risk-based premiums would help to alleviate the PPF's moral hazard problems. They argue that such premiums would simply drive already weak schemes and companies to insolvency: this is because for companies already close to insolvency, the correct premium would be close to the level of underfunding. And attempts to get more solvent schemes to cross-subsidise the weak schemes would exacerbate adverse selection and push the stronger schemes away from DB provision. Any attempt to make premiums risk-related could therefore easily backfire and increase the likelihood of the severe loss outcomes that the PPF is trying to avoid.

---

[16] This is precisely what Norwegian shipping group Aker Kvaerner did in the case of its UK subsidiary. Just days before the new pension regulatory regime came into force in April 2005, the subsidiary was sold to its management for £1, thereby breaking the link between a profitable parent and a loss-making subsidiary. At the time, the subsidiary had a pension scheme deficit of £245m on liabilities of £1.2bn. The Pension Regulator eventually pressurised Kvaerner into paying £101m into the fund before 2012, but this was still not enough to cover the full deficit.

**6.4 A funding standard for schemes**

There is also the possibility that moral hazard could be reduced by introducing tighter funding requirements. The 2004 Pensions Act recognised this possibility and introduced a *scheme-specific* funding standard (not a common (i.e., one size fits all) funding standard).

The Pensions Regulator (2006, para 2.4) announced it would take the following approach to implementing the requirements of the Act:
- to promote, through its code of practice and other forms of guidance and communication, good understanding by trustees, employers and their advisers of the matters they should consider when they agree their scheme's Statutory Funding Objective and any recovery plan needed to raise funding to that level;
- to intervene in those schemes where the funding objective is imprudent or the recovery plan is inappropriate, in order to protect members' benefits and/or reduce risks to the PPF; and
- to be transparent with trustees, employers and their advisers about the ways in which it intend to focus its resources on schemes that are likely to pose the greatest risk.

In doing this, it said it would use the following guiding principles (para 2.5):

- protect members – it would support trustees and employers working to maximise the protection of the benefits that the employer promised to pay and that members are expecting;
- be scheme specific – it is not its role, nor is it consistent with Government policy, to set a funding standard, because each scheme needs to take account of its particular circumstances;
- be risk-based – regulatory intervention should be focused on the schemes that pose the greatest risk to members' benefits and the PPF. While it is never possible to eliminate all risk, those in a position to do so should seek to mitigate those risks wherever it is reasonable to do so;
- be proportionate – trustees should aim to correct any shortfall as quickly as the employer can reasonably afford. The Pensions Regulator intends to distinguish between those schemes where rapid elimination of the shortfall would have a serious adverse impact on the employer's viability and those where employers could potentially afford to pay off the shortfall more quickly;
- be preventive – the Regulator needs where possible to act before risks materialise;

- be practicable – it needs need an approach that can be operated within the constraints of the information and resources available to it; and
- be a referee not a player – the responsibility for ensuring that schemes are fully funded rests with trustees and employers with the help of their advisers. The regulator will not interfere with this responsibility where it is properly discharged.

We would question whether the TPR's approach combined with the guiding principles will be sufficient to protect the PPF or whether a much more prescriptive funding standard is required, such as:

- a contribution rate set to remove a deficit over a short control period;
- a discount rate for determining the value of liabilities based on the risk-free rate to remove the possibility of the equity risk premium being used as an excuse to lower the contribution rate; or
- limits placed on the equity weighting in the pension fund.

The Pension Protection Fund (2006) has itself chosen a highly conservative investment strategy for the funds it receives when it takes over insolvent pension schemes:

| Asset class | Benchmark | Total percentage allocation |
|---|---|---|
| Cash | 3-Month LIBOR | 20% |
| Global bonds | J.P.Morgan Government Bond Index | 50% |
| UK equities | FTSE All-Share | 12.5% |
| Global equities | FTSE Global | 7.5% |
| Property | IPD | 7.5% |
| Currency overlay | 3-Month LIBOR | 2.5% |

The Pension Protection Fund (2006, paras 5.2.3 and 5.2.5) explains this choice as follows: the asset allocation has been set after maximising the expected excess return over the liability subject to the following constraints:

- A risk budget of 4% p.a. at the total fund level. The risk budget is the maximum ex-ante standard deviation of the difference between the asset return and the return on the 'liability benchmark'. This liability benchmark is the notional portfolio of assets that exactly matches the expected liability cashflows.
- Each asset class is actively managed with tracking error limits and out-performance targets such that the contribution of expected active manager excess returns to total out-performance is a maximum of 25%. Tracking error is the amount of divergence of the performance of the fund against the specified benchmark.

- A portfolio of derivatives known as a 'swap overlay' is applied to the portfolio above to change its cash flow profile to match that of the 'liability benchmark'. This ensures that the sensitivities to real and nominal interest rates of the asset values closely match that of the liabilities.

Given the investment strategy and the 4% total risk budget of 4%, the PPF expects the return on investments to exceed the return on the liability benchmark by 1.4% p.a.

However, the PPF has made it clear that its investment strategy should not be taken as a blueprint by other pension funds. The investment strategy has been criticised for being too conservative: the consequence of foregoing the equity risk premium on 80% of its investments is likely to be higher contributions to the PPF levies according to critics, once again emphasing the tension in the tradeoff between the level of contributions on the one hand and the riskiness of the assets purchased with those contributions on the other (*Pensions Week*, 12 October 2006).

## 6.5 Close supervision and the public exposure of companies that underfund their schemes

Schemes with large deficits need to be watched and supervised very closely, and in this the PPF can learn a lot from the PBGC's Early Warning Program. However, it is far from enough for the PPF merely to watch and supervise 'problem' funds: it also needs to provide incentives for sponsors to take their responsibilities seriously.

One promising way to provide such an incentive is provided by a clause in the 2004 Pensions Act that allows the Pensions Regulator to issue a 'contribution notice' (CN) requiring a person who has been involved (within the previous six years) in a deliberate act to avoid pension liabilities to put money into a pension scheme up to a specified amount or to issue a 'financial support direction' (FSD) requiring associated or connected persons to put financial support in place to guarantee the pension liabilities of an insufficiently resourced sponsor. A 'clearance procedure' with the Pensions Regulator can be used to ensure that actions (called type A events) will not lead to the issue of a CN or FSD for schemes in deficit. Examples of type A events are the payment of a large dividend, a large sale buyback, or the sale of the firm to another highly leveraged firm. These anti-avoidance powers are 'unprecedented in the history of company law and the lifting of the corporate veil should send a shiver down the spine of all irresponsible directors, their advisers and professional indemnity insurers' (Farr (2005, p. 21)). According to Farr, this should give comfort to trustees who are likely to be the largest material unsecured creditors of the sponsoring firm. At the same time, Farr suggests that these trustees should also seek the advice of specialists in creditor negotiations.

## 7 Conclusions

Unlike banks and insurance companies, UK pension funds are not regulated by the FSA and, moreover, are not subject to formal capital requirements. Instead they operate on a prudent person principle and make promises not guarantees. This goes a long way towards explaining why the current financial regulation of pension funds is so different from that of banks and life assurers. And, yet, as we have seen, it is widely expected that pension funds should actually deliver the pension outcomes they promise, and in any case the new PPF itself is in the business of providing guarantees. Indeed, one can argue that the establishment of the PPF has radically altered the nature of the game by turning the promise of the scheme sponsor into a guarantee by the PPF that is (arguably, though the Government denies this) underwritten ultimately by the taxpayer.

This raises some deep and difficult issues for both the regulation of the pension fund industry and for the PPF. One might argue, for example, that the similarities between pension funds and other financial institutions – life assurers especially – are so strong and compelling that it makes no sense to subject them to such different regulatory regimes. There is therefore an important issue of harmonisation (or lack thereof) between the capital regulation regimes applied to pension funds and those applied other financial institutions. Of course, our discussion does not address which of these regulatory regimes might be best, or whether they should all be replaced in favour of some other regime. If one accepts that the three-pillar approach is broadly 'right', at least in principle – as many do – then one would be tempted to suggest that it should be extended to cover pension funds too. On the other hand, many features of the three-pillar approach have been extensively and heavily criticised – such as its cumbersome inflexibility, its emphasis on the VaR risk measure[17], its dependence on ad hoc assumptions, and so on, and even the principle of risk-based capital requirements is open to dispute (Danielsson (2003)). Good arguments can therefore be made against the whole approach. If one agrees with this line of reasoning, extending the three-pillar approach to pensions might make little sense, even if it did help to harmonize the regulatory treatment of pension funds and other financial institutions.

There are also related issues arising from the very diverse ways in which different types of financial institution measure and manage their financial risks. The different types of financial institution have different approaches to risk management and vary considerably in their degrees of risk management sophistication. There is a general perception that the most 'advanced' risk management practices are to be found in capital markets institutions and banks. Insurance companies are generally perceived to be less sophisticated, although the better reinsurance companies would appear to be as good at risk management as any capital market institution or bank. For their part, pension funds are generally perceived to be backward in their risk management practices and we would share this assessment. Pension funds have a lot to learn about risk management. This is an especially important point when one also considers that pension fund risk management is an inherently complex matter: there are difficult valuation problems, complex embedded options, tricky risk factors (e.g., mortality), and so on.

Yet different types of financial institution deal with different problems – they vary considerably in the types of risk they face, the horizons they operate to, the inherent complexity of the portfolios they handle, the liquidity of the markets in which they operate, and in many other ways besides – and it should go without saying that any transfer of risk management technology or practice into the pensions sector needs to take account of the uniqueness of the institutional contexts in which pension funds operate. These differences between different sectors of the financial services industry might also affect the ways in which these sectors can be regulated: what is feasible in one sector might not be feasible in another.

---

[17] For a critique of VaR, see Dowd and Blake (2006).

Turning now to the PPF, we would argue that the PPF faces a daunting uphill struggle. It is essentially offering put options on pension scheme assets, but these options are highly complex and very hard to price.

This said, the PPF would certainly help itself if it learned from the experience of its (not too successful) counterpart in the US, the PBGC: the PBGC does make a serious effort to monitor problem funds and anticipate undesirable outcomes, and many of its risk management practices – its Early Warning Program, and so on – are tried and tested, and would no doubt be useful to the PPF. But learning takes time and time may not be on its side. We would not rule out the possibility that one or two major 'hits' – a failure of a couple of firms the size of British Airways, for example – might bring it down much more quickly than anyone expected. And even if it manages to avoid this fate, its chances of surviving a major recession cannot be considered high.

Perhaps the root problem is that the PPF has been established on a contradictory foundation. On the one hand, the Government insists that the PPF will provide pension guarantees, but on the other hand, the Government also insists that the PPF should be 'on its own' and not expect any Government bailout if it gets into difficulties. We would argue that the Government's position is contradictory, because the PPF has only a restricted capacity to protect its own solvency. Any 'guarantees' it provides are therefore inevitably limited ones, and the Government is reluctant to face up to the reality that this implies. We are *not* suggesting that the Government should underwrite the PPF – far from it – but we are only pointing out that the Government has not thought its position through. The inevitable consequence of this position is that the costs of financial regulation will increase as both the Pensions Regulator and the Pension Protection Fund attempt to protect themselves as the system gets into difficulties.[18]

This lack of 'joined up thinking' is also illustrated by the fact that the very Act of Parliament that established the PPF also replaced the MFR with a much weaker scheme-specific funding standard, i.e., the Act put more weight on the funding standard whilst simultaneously weakening it, and yet the funding standard was not strong to begin with.

To end on a really gloomy note, consider the following: Since the average FRS17 funding level for the UK's top 350 companies is about 80%, these companies would need an extra £150bn to cover the PPF level of funding (*Pensions Week*, 25 April 2005). This indicates that in its first month after launch, the PPF was providing in excess of £150bn of insurance cover against annual levy premiums of just £300m. However, Standard & Poor's carried out a study in 2005 of potential claims against the PPF, based on the post-1981 default experience of 340 top UK companies, and their results suggest that annual claims on the PPF will exceed £300m. Even under the most optimistic assumptions, under which the PPF would recover 40% of the deficit from the defaulted sponsor, the annual claim on the PPF would be £670m. This does not look good. And, in the first month alone of the PPF's existence, it received claims in excess of £1bn with the collapse of Turner & Newell (liabilities of £875m) and Rover (liabilities of £400m). If this is a good start, one wonders what a bad one would look like.

---

[18] At the time of writing this article, these were the type of headlines that were common: 'Funding regs trigger higher costs' (Pensions Week, 30 October 2006) and 'Pensions safety net levy rises by 50%' (Daily Telegraph, 28 October 2006).

Our overall conclusions are therefore deeply disturbing: right from its birth, the PPF faces the permanent risk of insolvency as a consequence of the moral hazard, adverse selection, and, especially, systemic risks that it faces. If this unfortunate eventuality happens, the Government will then face the unpleasant choice of whether to bail out the PPF or allow a mandatory Government-sponsored insurance system to collapse.